\documentclass[nocrop]{bioinfo} 
\copyrightyear{2023}
\pubyear{2023}

\usepackage{amsmath}
\usepackage{amsthm}
\usepackage{amssymb}
\usepackage{graphicx}
\usepackage[colorlinks=true,linkcolor=red,urlcolor=blue,citecolor=magenta,allcolors=blue]{hyperref}
\usepackage{tablefootnote}
\usepackage{multirow}
\usepackage{booktabs}
\usepackage{float}
\usepackage[noend]{algpseudocode}
\usepackage[ruled]{algorithm}
\newtheorem{thm}{Theorem}
\newtheorem{lem}{Lemma}

\newtheorem{prop}{Property}
\newtheorem{defn}{Definition}

\newtheorem{remark}{Remark}

\DeclareMathSymbol{\mlqq}{\mathrel}{operators}{"5C}
\DeclareMathSymbol{\mrqq}{\mathrel}{operators}{`"}

\newcommand{\return}{\vspace{0.125cm}}
\newcommand{\parag}[1]{\vspace{0.2cm}\textbf{{#1.}}}

\newcommand{\afterfigspace}{\vspace{-3mm}}

\newcommand{\func}[1]{\textsf{\mbox{#1}}}
\newcommand{\pr}[1]{\mathbb{P}{[#1]}}
\newcommand{\txt}[1]{\textup{#1}}

\newcommand{\code}[1]{\mbox{\textbf{#1}}}
\newcommand{\var}[1]{\mbox{\emph{#1}}}

\usepackage{xcolor, soul}

\newcommand{\red}[1]{{\color{black}{#1}}}

\newcommand{\method}[1]{\mbox{\textup{{#1}}}}

\newcommand{\nuc}[1]{\method{\textsf{#1}}}
\newcommand{\kmer}{$k$-mer}
\newcommand{\skm}{super-{\kmer}}

\newcommand{\spectrum}{\mbox{\textup{spectrum}}}

\newcommand{\celeg}{\method{Elegans}}
\newcommand{\cod}{\method{Cod}}
\newcommand{\kes}{\method{Kestrel}}
\newcommand{\human}{\method{Human}}

\newcommand{\yeast}{\method{Yeast}}

\newcommand{\Next}{\method{Next}}

\newcommand{\Access}{\method{Access}}

\newcommand{\Rank}{\method{Rank}}

\access{}
\appnotes{}

\begin{document}
\firstpage{1}


\title[Locality-Preserving Minimal Perfect Hashing of K-Mers]{Locality-Preserving Minimal Perfect Hashing of K-Mers}

\author[G. E. Pibiri, Y. Shibuya, and A. Limasset]{Giulio Ermanno Pibiri\,$^{\text{\sfb 1,2}}$, Yoshihiro Shibuya\,$^{\text{\sfb 3}}$, Antoine Limasset\,$^{\text{\sfb 4}}$}

\address{%
$^{\text{\sf 1}}$Ca' Foscari University of Venice, Venice, Italy\\
$^{\text{\sf 2}}$ISTI-CNR, Pisa, Italy\\
$^{\text{\sf 3}}$University Gustave Eiffel, Marne-la-Vall\'ee, France\\
$^{\text{\sf 4}}$University of Lille and CNRS, Lille, France%
}

\corresp{}
\history{}
\editor{}


\abstract{\textbf{Motivation:} Minimal perfect hashing is the problem of mapping a static set of $n$
distinct keys into the address space $\{1,\ldots,n\}$ bijectively.
It is well-known that
$n\log_2(e)$ bits are necessary to specify a minimal perfect hash function (MPHF) $f$,
when \emph{no additional knowledge} of the input keys is to be used.
However, it is often the case in practice that the input keys
have intrinsic relationships that we can exploit to lower the bit complexity
of $f$.
For example, consider a string and the set of all its
distinct {\kmer}s as input keys:
since two consecutive {\kmer}s share an overlap of $k-1$ symbols,
it seems possible to beat the classic $\log_2(e)$ bits/key barrier
in this case.
Moreover, we would like $f$ to map consecutive {\kmer}s
to consecutive addresses, as to also \emph{preserve} as much as possible
their relationship in the codomain.
This is a useful feature in practice
as it guarantees a certain degree of \emph{locality} of reference for $f$,
resulting in a better evaluation time when querying consecutive {\kmer}s.
\\
\noindent
\textbf{Results:} Motivated by these premises, we initiate the study of a new type of
locality-preserving MPHF designed for {\kmer}s
extracted consecutively from a collection of strings.
We design a construction whose space usage decreases
for growing $k$ and discuss experiments with a practical implementation of the method:
in practice, the functions built with our method can be several times smaller
and even faster to query than the most efficient MPHFs in the literature.
%
\\
\noindent
\textbf{Code Availability:}
\url{https://github.com/jermp/lphash}
%
\\
\noindent
\textbf{Data Availability:}
\url{https://zenodo.org/record/7239205}}

\maketitle

\section{Introduction}\label{sec:introduction}

Given a universe set $U$,
a function $f : U \rightarrow [n] = \{1,\ldots,n\}$
is a \emph{minimal perfect hash function} (MPHF, henceforth)
for a set $S \subseteq U$
with $n=|S|$ if $f(x) \neq f(y)$ for all $x,y \in S$, $x \neq y$.
In simpler words, $f$ maps each key of $S$ into a distinct integer
in $[n]$. The function is allowed to return any value in $[n]$
for a key $x \in U \setminus S$.
A classic result established that
$n\log_2(e) = 1.442n$ bits are essentially necessary
to represent such functions for $|U| \gg n$~\citep{mehlhorn1982program}. 
Minimal perfect hashing is a central problem in data structure
design and has received considerable attention, both in theory
and practice.
In fact, many practical constructions have been proposed
(see, e.g.,~\citep{PibiriT21} and references therein).
These algorithms find MPHFs that take space close to the theoretic-minimum, e.g.,
2 -- 3 bits/key, retain very fast lookup time,
and scale well to very large sets.
Applications of minimal perfect hashing range from
computer networks~\citep{lu2006perfect}
to databases~\citep{chang2005perfect},
as well as 
language models~\citep{PibiriV19,StrimelRTPW20}, 
compilers, and operating systems.
MPHFs have been also used recently in Bioinformatics
to implement fast and compact dictionaries for fixed-length
DNA strings~\citep{pibiri2022sshash,pibiri2022WABI,almodaresi2018space,blight}.

In its simplicity and versatility, the minimal perfect hashing problem
does not take into account specific types of inputs, nor the
intrinsic relationships
between the input keys.
Each key $x \in S$ is considered independently
from any other key in the set and,
as such, $\pr{f(x)=i} \approx \frac{1}{n}$ for any fixed $i \in [n]$.
In practice, however, the input keys often
present some regularities that we could exploit to let $f$ act
``less randomly'' on $S$. This, in turn, would permit to achieve
a \emph{lower} space complexity for $f$.

We therefore consider in this paper the following
special setting of the minimal perfect hashing problem:
the elements of $S$ are all the distinct sub-strings of length $k$, for some $k>0$,
from a given collection $\mathcal{X}$ of strings.
The elements of $S$ are called {\kmer}s.
The crucial point is that any two consecutive {\kmer}s in a string of $\mathcal{X}$
have indeed a strong intrinsic
relationship in that they share an overlap of $k-1$ symbols.
It seems profitable to exploit the overlap information to
\emph{preserve} (as much as possible) the \emph{local} relationship
between consecutive {\kmer}s
as to \emph{reduce} the randomness of $f$, thus lowering its bit complexity
and evaluation time.

In particular, we are interested in the design of a \emph{locality-preserving} MPHF
in the following sense.
Given a query sequence $Q$, if $f(x)=j$ for some {\kmer} $x \in Q$,
we would like $f$ to hash $\Next(x)$ to $j+1$,
$\Next(\Next(x))$ to $j+2$, and so on, where $\Next(x)$ is the {\kmer}
following $x$ in $Q$ (assuming $\Next(x)$ and $\Next(\Next(x))$ are in $\mathcal{X}$ as well).
This behavior of $f$ is very desirable in practice, at least for two important
reasons.
First, it implies \emph{compression} for satellite values associated to {\kmer}s.
Typical satellite values are abundance counts, reference identifiers
(sometimes called ``colors''),
or contig identifiers (e.g., unitigs) in a de Bruijn graph.
Consecutive {\kmer}s tend to have very similar --
if not identical -- satellite values, hence hashing consecutive {\kmer}s to
consecutive identifiers induce a natural clustering of the associated satellite values
which is amenable to effective compression.
The second important reason is, clearly, faster evaluation time
when querying for consecutive {\kmer}s in a sequence.
This \emph{streaming} query modality is the query modality employed by
{\kmer}-based applications~\citep{almodaresi2018space,bingmann2019cobs,blight,findere,pibiri2022sshash}.


We formalize the notion of locality-preserving MPHF along with other
preliminary definitions in Section~\ref{sec:definitions}.
We show how to obtain a locality-preserving MPHF in very compact space
in Section~\ref{sec:upper-bound}.
To achieve this result, we make use of two algorithmic tools:
\emph{random minimizers}~\citep{schleimer2003winnowing,roberts2004reducing}
and a novel partitioning scheme for sub-sequences of consecutive {\kmer}s
sharing the same minimizers (super-{\kmer}s) which allows
a more parsimonious memory layout.
The space of the proposed solution decreases for growing $k$
and the data structure is built in linear time in the size of the input
(number of distinct {\kmer}s).
In Section~\ref{sec:experiments} we present experiments
across a breadth of datasets
to show that the construction is practical too:
the functions can be several times smaller
and even faster to query than the most efficient, albeit ``general-purpose'',
minimal perfect hash functions.
We conclude in Section~\ref{sec:conclusion} where we also sketch some
promising future directions. 
Our C++ implementation of the method is publicly available at
\url{https://github.com/jermp/lphash}.

\section{Notation and Definitions}\label{sec:definitions}

Let $\mathcal{X}$ be a set of strings over an alphabet $\Sigma$.
Throughout the paper
we focus on the DNA alphabet $\Sigma=\{\nuc{A},\nuc{C},\nuc{G},\nuc{T}\}$
to better highlight the connection with our concrete application
but our algorithms
can be generalized to work for arbitrary alphabets.
A sub-string of length $k$ of a string $S \in \mathcal{X}$
is called a {\kmer} of $S$.


\begin{defn}[Spectrum]
The {\kmer} spectrum of $\mathcal{X}$
is the set of all distinct {\kmer}s
of the strings in $\mathcal{X}$. Formally:
$
\spectrum_k(\mathcal{X}) := \{ x \in \Sigma^k \,\, | \,\, \exists S \in \mathcal{X} \text{ such that } x \text{ is a {\kmer} of } S \}.
$
\end{defn}

\begin{defn}[Spectrum-Preserving String Set]
A spectrum-preserving string set (or SPSS) $\mathcal{S}$
of $\mathcal{X}$
is a set of strings such that (i)
each string of $\mathcal{S}$ has length at least $k$, and (ii)
$\spectrum_k(\mathcal{S})=\spectrum_k(\mathcal{X})$.
\end{defn}

Since our goal is to build a MPHF for the {\kmer}s in a SPSS,
we are interested in a SPSS $\mathcal{S}$ where
each {\kmer} is seen only once, i.e., for
each {\kmer} $x \in \spectrum_k(\mathcal{S})$ there is only
one string of $\mathcal{S}$ where $x$ appears once.
We assume that no {\kmer} appearing at the end of a string
shares an overlap of $k-1$ symbols with the first {\kmer} of another
string,
otherwise we could reduce the number of strings in $\mathcal{S}$
and obtain a smaller SPSS.
In the following, we make use of this form of SPSS which is suitable for
the minimal perfect hashing problem.
We remark that efficient algorithms exist
to compute such SPSSs (see, e.g.,~\citep{rahman2020representation,bvrinda2021simplitigs,khan2021cuttlefish,khan2022scalable}).

The input for our problem
is therefore a SPSS $\mathcal{S}$ for $\mathcal{X}$ with $|\mathcal{S}|$ strings
and $n>1$ distinct {\kmer}s.
Without loss of generality, we index {\kmer}s based on their positions in $\mathcal{S}$,
assuming an order
$S_1,S_2,S_3,\ldots$ of the strings of $\mathcal{S}$ is fixed,
and we indicate with $x_i$ the $i$-th {\kmer} in $\mathcal{S}$,
for $i=1,\ldots,n$.

We want to build a MPHF $f : \Sigma^k \rightarrow [n]$
for $\mathcal{S}$;
more precisely,
for the $n$ distinct {\kmer}s in $\spectrum_k(\mathcal{S})$.
We remark again that our objective is to exploit
the overlap of $k-1$ symbols between consecutive {\kmer}s from a string
of $\mathcal{S}$
to preserve their locality,
and \emph{hence} reduce the bit complexity of $f$
as well as its evaluation time when querying {\kmer}s in sequence.

\return
We define a \emph{locality-preserving} MPHF, or LP-MPHF, for $\mathcal{S}$ as follows.

\begin{defn}[LP-MPHF]
\label{def:lpmphf}
Let $f : \Sigma^k \rightarrow [n]$ be a MPHF for $\mathcal{S}$
and $A$ be the set
$\{ 1 \leq i < n \, | \, \exists S \in \mathcal{S}, x_i, x_{i+1} \in S \wedge f(x_{i+1})=f(x_i)+1\}$.
The function $f$
is $(1-\varepsilon)$-locality-preserving for $\mathcal{S}$
if \red{$\varepsilon \geq 1 - |A|/n$}.
\end{defn}

Intuitively, the ``best'' LP-MPHF for $\mathcal{S}$ is the one having the smallest
$\varepsilon$, so we look for practical constructions with small $\varepsilon$.
On the other hand,
note that a ``classic'' MPHF corresponds to the case where
the locality-preserving property is almost always \emph{not} satisfied
and, as a consequence, $\varepsilon$ will be approximately 1.

Two more considerations are in order.
First, it should be clear that the way we define locality-preservation in Definition~\ref{def:lpmphf}
is only pertinent to SPSSs where
having consecutive hash codes for consecutive {\kmer}s
is a very desirable property as motivated in Section~\ref{sec:introduction}.
A different definition of locality-preservation could instead be given
if we were considering generic input keys.
Second, we did not use the term \emph{order-preserving} to stress the
distinction from classic order-preserving functions in the literature~\citep{fox1991order}
that make it possible to preserve \emph{any} wanted order and, as such, incur in
an avoidable $\Omega(\log n)$-bit overhead per key.
Here, we are interested in preserving only the input order of the {\kmer}s
which is the one that matters in practice.

\begin{defn}[Fragmentation Factor]\label{def:frag-fact}
Given a SPSS $\mathcal{S}$ with $|\mathcal{S}|$ strings
and $n=|\spectrum_k(\mathcal{S})|$ distinct {\kmer}s,
we define the fragmentation factor of $\mathcal{S}$
as $\alpha := (|\mathcal{S}|-1)/n$.
\end{defn}

The fragmentation factor of $\mathcal{S}$ is a measure of how contiguous
the {\kmer}s in $\mathcal{S}$ are.
The \emph{minimum} fragmentation $\alpha=0$ is achieved for $|\mathcal{S}|=1$
and, in this case, $x_i$ shares an overlap of $k-1$ symbols with $x_{i+1}$
for \emph{all} $i=1,\ldots,n-1$.
This ideal scenario is, however, unlikely to happen in practice.
On the other hand, the worst-case scenario of maximum fragmentation
$\alpha=1-1/n$ is achieved when $|\mathcal{S}|=n$
and {\kmer}s do not share any overlap (of length $k-1$).
This is also unlikely to happen given that {\kmer}s are extracted consecutively
from the strings of $\mathcal{X}$ and, as a result, many overlaps are expected.
A more realistic scenario happens, instead, when
$|\mathcal{S}| \ll n$, resulting in $\varepsilon \gg \alpha$.
For the rest of the paper, we focus on this latter scenario
to make our analysis meaningful.

From Definition~\ref{def:lpmphf} and~\ref{def:frag-fact}
it is easy to see that $\varepsilon \geq 1/n$ when $\alpha=0$,
and $\varepsilon=1$ when $\alpha=1-1/n$. In general,
we have $\varepsilon \geq \alpha + 1/n$ since there are at least
$|S|-1$ indexes $i$ for which $f(x_{i+1}) \neq f(x_i)+1$.
How small $\varepsilon$ can actually be therefore depends on the input SPSS
(and on the strategy used to implement $f$ in practice, as we are going to
illustrate in Section~\ref{sec:upper-bound}).

\return
Lastly in this section, we define
minimizers and {\skm}s that will be one of the main ingredients
used in Section~\ref{sec:upper-bound}.

\begin{defn}[Random Minimizer of a {\kmer}]\label{def:random-minimizer}
Given a {\kmer} $x$ and a random hash function $h$,
the minimizer of $x$ is any $m$-mer $\mu$ such that $h(\mu) \leq h(y)$
for any other $m$-mer $y$ of $x$,
for some $m \leq k$.
\end{defn}

In case the minimizer of $x$ is not unique, we break ties by taking
the leftmost $m$-mer in $x$.
For convenience, we indicate with $w=k-m+1$ the number
of $m$-mers in a {\kmer}. (Note that Definition~\ref{def:random-minimizer}
defines a minimizer
as a specific $m$-mer inside a {\kmer} rather than a specific {\kmer} in a window
of $w$ consecutive {\kmer}s, which is the more standard definition found in the literature.)
Since $h$ is a random hash function (with a wide range, e.g., $[1..2^{64}]$),
each $m$-mer in a {\kmer} has probability $\approx\frac{1}{w}$ of being the minimizer of the {\kmer}.
We say that the triple $(k,m,h)$ defines a random minimizer \emph{scheme}.
The \emph{density} of a minimizer scheme is the expected number of
selected minimizers from the input.

\begin{defn}[Super-{\kmer}]
\red{Given a string $S$, a {\skm} $g$ is a maximal sub-string of $S$ where each {\kmer}
has the same minimizer $\mu$ and $\mu$ appears only once in $g$.}
\end{defn}

\raggedbottom

\section{Locality-Preserving Minimal Perfect Hashing of K-Mers}\label{sec:upper-bound}

In this section we describe an algorithm to obtain
locality-preserving MPHFs for a spectrum-preserving string set $\mathcal{S}$.
The algorithm
builds upon the following main insight.

\begin{figure}[t]
    \centering
    \includegraphics[scale=0.6,type=pdf,ext=.pdf,read=.pdf]{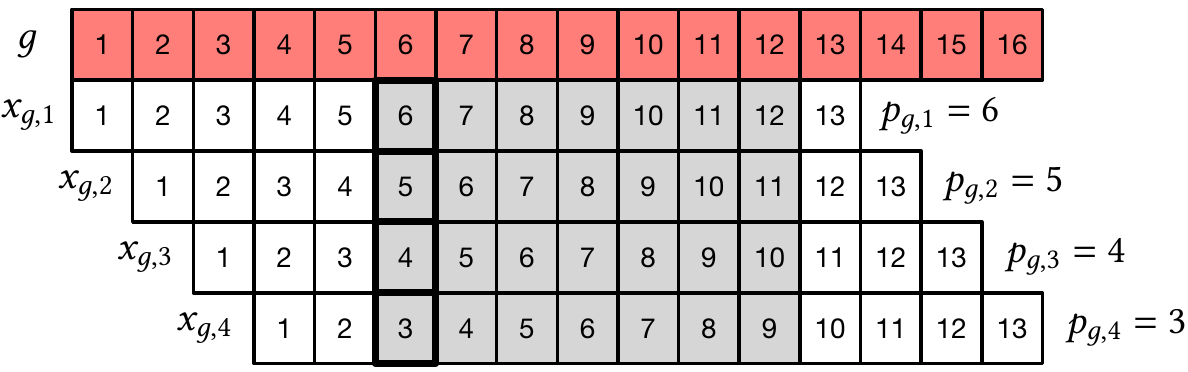}
    \vspace{-4mm}
    \caption{A {\skm} $g$ of length 16 with $|g|-k+1=16-13+1=4$ {\kmer}s
    $x_{g,1},x_{g,2},x_{g,3},x_{g,4}$ for $k=13$
    and minimizer length $m=7$. The shaded boxes highlight the minimizer
    whose start position is $p_{g,i}$ in {\kmer} $x_{g,i}$.
    It is easy to see that $i=p_{g,1}-p_{g,i}+1$ for any $1 \leq i \leq |g|-k+1$.
    }
    \label{fig:skm}
    {\afterfigspace}
\end{figure}

\parag{Implicitly Ranking {\kmer}s through Minimizers}
Let $g$ be a {\skm} of some string $S\in\mathcal{S}$
and assume $g$ is the only {\skm} whose minimizer is $\mu$.
By definition of {\skm},
all the {\kmer}s $x_{g,1},\ldots,x_{g,|g|-k+1}$ in $g$ contain the minimizer $\mu$
as a sub-string -- $x_{g,i}$ being the $i$-th {\kmer} of $g$.
If $p_{g,1}$ is the start position of $\mu$ in the first {\kmer} $x_{g,1}$ of $g$,
then
\begin{equation}\label{eq:pos}
p_{g,i}=p_{g,1}-i+1
\end{equation}
is the start position of $\mu$ in $x_{g,i}$
for $1 \leq i \leq |g|-k+1$.
Fig.~\ref{fig:skm} gives a practical example for a {\skm} $g$
of length 16 and $k=13$.

The next property illustrates the relation between the size $|g|-k+1$ of the {\skm} $g$
and the position $p_{g,1}$ (we will come later on the implications of this property).

\begin{prop}\label{prop:relation-size-pos}
$|g|-k+1 \leq p_{g,1} \leq w$ for any {\skm} $g$.
\end{prop}

\begin{proof}
Since $p_{g,1}$ is the start position of the minimizer in the first {\kmer} of $g$,
there are at most $p_{g,1}$ {\kmer}s that contain the minimizer as a sub-string,
hence $|g|-k+1 \leq p_{g,1}$. However, $g$ cannot contain more than $w$ {\kmer}s.
\qedsymbol
\end{proof}

Now, suppose we are given a query {\kmer} $x \in \mathcal{S}$ whose minimizer is $\mu$.
The {\kmer} must appear as a sub-string of $g$, i.e., it must be
one among $x_{g,1},\ldots,x_{g,|g|-k+1}$.
We want to compute the rank of $x$
among the {\kmer}s $x_{g,1},\ldots,x_{g,|g|-k+1}$ of $g$, which we indicate by $\Rank(x)$
(assuming that it is clear from the context that $\Rank$ is relative to $g$).
Let $p$ be the start position of $\mu$ in $x$.
We can use this positional information $p$ to compute $\Rank(x)$
as follows:
\begin{itemize}
    \item if $p_{g,1} \geq p$ and $1 \leq p_{g,1}-p+1 \leq |g|-k+1$, then
    \begin{equation}\label{eq:local-rank}
    \Rank(x) = p_{g,1}-p+1
    \end{equation}
    \item otherwise ($p_{g,1} < p$ or $p_{g,1}-p+1 > |g|-k+1$),
    $x$ cannot possibly be
    in $g$ and, hence, indexed by $f$.
\end{itemize}

Our strategy is to compute $f(x_{g,i})$ as
\begin{equation}\label{eq:strategy}
f(x_{g,i}) = f(x_{g,1}) + \Rank(x_{g,i}) - 1 = f(x_{g,1}) + p_{g,1} - p_{g,i}
\end{equation}
for any {\kmer} $x_{g,1},\ldots,x_{g,|g|-k+1}$ of $g$.
Next, we show in Lemma~\ref{lem:local-rank} that this strategy
maps the {\kmer}s $x_{g,1},\ldots,x_{g,|g|-k+1}$ bijectively in
$\{(f(x_{g,1})-1) + 1, \ldots, (f(x_{g,1})-1) + |g|-k+1\}$
\emph{and preserves their locality} (i.e., their relative order in $g$).

\begin{lem}\label{lem:local-rank}
The strategy in Equation~\ref{eq:strategy} guarantees
$f(x_{g,i+1})=f(x_{g,i})+1$ for any $i=1,\ldots,|g|-k$.
\end{lem}
\begin{proof}
For Equation~\ref{eq:strategy},
$f(x_{g,i})=f(x_{g,1})+p_{g,1}-p_{g,i}$.
Therefore $f(x_{g,i+1})=f(x_{g,1})+p_{g,1}-p_{g,i+1}$.
Since $p_{g,i+1}=p_{g,i}-1$ for Equation~\ref{eq:pos}, then
$f(x_{g,i+1})=f(x_{g,1})+p_{g,1}-p_{g,i+1}=f(x_{g,1})+p_{g,1}-p_{g,i}+1=f(x_{g,i})+1$.
\qedsymbol
\end{proof}

To sum up,
the position of the minimizer
in the first {\kmer} of $g$, $p_{g,1}$,
defines an \emph{implicit ranking}
(i.e., achieved without explicit string comparison)
of the {\kmer}s inside a {\skm}.

\subsection{Basic Data Structure}\label{sec:data-structure}

From Equation~\ref{eq:strategy} is evident
that $f(x_{g,1})$ acts as a ``global'' component in the calculation of $f(x_{g,i})$,
which must be added to a ``local'' component represented by $\Rank(x_{g,i})$.
We have already shown how to compute $\Rank(x_{g,i})$ in Equation~\ref{eq:local-rank}:
Lemma~\ref{lem:local-rank} guarantees that this local rank computation
bijectively maps the {\kmer}s of $g$ into $[1..|g|-k+1]$.
We are therefore left to show how to compute $f(x_{g,1})$ for each {\skm} $g$.
We proceed as follows.

\begin{algorithm}[t]
\begin{algorithmic}[1]
\Function{$f$}{$x$}:
    \State{$(\mu,p) = \func{minimizer}(x)$}
    \State{$i=f_m(\mu)$}
    \State{\code{return} $L[i] + P[i] - p$}
\EndFunction
\end{algorithmic}
\caption{Evaluation algorithm for $f$, given the {\kmer} $x$.
The helper function
$\func{minimizer}(x)$ computes the minimizer $\mu$ of $x$
and the starting position $p$ of $\mu$ in $x$.
\label{alg:lookup-v1}}
\end{algorithm}


\parag{Layout}
Let $\mathcal{M}$ be the set of all the distinct minimizers of $\mathcal{S}$.
We build a MPHF for $\mathcal{M}$, $f_m : \Sigma^m \rightarrow [|\mathcal{M}|]$.
Assume, for ease of exposition, that each {\skm} $g$
is the only {\skm} having minimizer $\mu$.
(We explain how to handle the case where more {\skm}s have the same
minimizer in Section~\ref{sec:ambiguous-min}.)
\red{We allocate an array $L'[1..|\mathcal{M}|+1]$ where $L'[1]=0$
and $L'[f_m(\mu)+1]=|g|-k+1$ for every minimizer $\mu$.
We then take the prefix-sums of $L'$ into another array $L$,
that is, $L[i]=\sum_{j=1}^i L'[j]$ for all $i=2,\ldots,|\mathcal{M}|+1$.
We therefore have that $L[f_m(\mu)]$ indicates the number of {\kmer}s
before those in $g$ (whose minimizer is $\mu$) in the order given by $f_m$.}
The size of $g$ can be recovered as
$L[f_m(\mu)+1]-L[f_m(\mu)]=|g|-k+1$.
In conclusion, we compute $f(x_{g,1})$ as $L[f_m(\mu)]$.
The positions $p_1$ of each {\skm} $g$ are instead written in another array
$P[1..|\mathcal{M}|]$ where $P[f_m(\mu)]=p_1$.

It follows that the data structure is built in $O(n)$ time, since a scan
over the input suffices to compute all {\skm}s and $f_m$ can be built
in $O(|\mathcal{M}|)$ expected time.

\parag{Lookup}
With these three components -- $f_m$, and the two arrays $L$ and $P$ --
it is easy to evaluate $f(x)$ as shown in Algorithm~\ref{alg:lookup-v1}.
The complexity of the lookup algorithm is $O(w)$ since this is the complexity
of computing the minimizer (assuming each hash calculation to take constant time)
and the overall evaluation of $f_m$ as well,
since accessing the arrays $L$ and $P$ takes $O(1)$.


\begin{figure}[t]
    \centering
    \includegraphics[scale=0.55,type=pdf,ext=.pdf,read=.pdf]{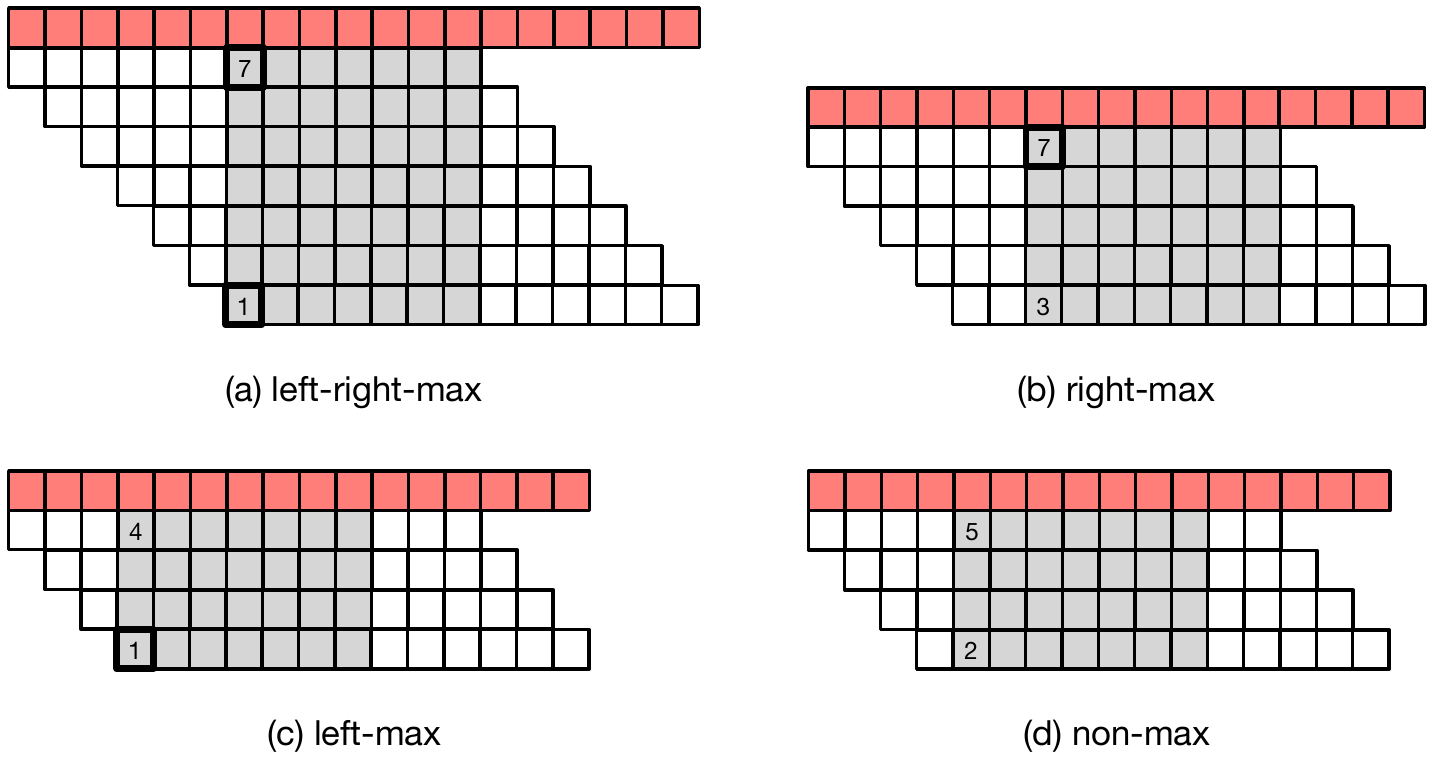}
    \vspace{-5mm}
    \caption{The four different types of {\skm}s. The example is
    for $k=13$ and minimizer length $m=7$, so $w=k-m+1=13-7+1=7$.
    The shaded boxes highlight the minimizer sub-string inside a {\kmer}.
    The start position of the minimizer is marked with a solid border when it is either max (7),
    or min (1).}
    \label{fig:skms}
\end{figure}

\parag{Compression}
The data structure for $f$ itself is a compressed representation for $f_m$, $L$, and $P$.
To compute the space taken by the data structure we first need
to know $|\mathcal{M}|$ -- the expected number of distinct minimizers seen in the input.
Assuming again that there are no duplicate minimizers,
if $d$ indicates the density of a random minimizer scheme, then
\begin{itemize}
    \item $|\mathcal{M}|=dn$, and
    \item $\varepsilon = d$ \red{as a direct consequence of} Lemma~\ref{lem:local-rank}.
\end{itemize}
In particular, a
result due to Zheng et al.~\citep[Theorem 3]{zheng2020improved} allows us to compute
$d$ for a random minimizer scheme as $d=\frac{2}{w+1}+o(1/w)$
if $m > (3+\epsilon)\log_4(w+1)$ for any $\epsilon>0$.
We will always operate under the condition that
$m$ is sufficiently large compared to $k$ otherwise minimizers
are meaningless.

Therefore any random minimizer scheme gives us a $(1-\varepsilon)$-LP MPHF
with $\varepsilon=\frac{2}{w+1}$ (we omit lower order terms for simplicity)
as illustrated in the following theorem
(see the Supplementary material for the proof).

\begin{thm}\label{thm:space-usage-random-minimizers-1}
Given a random minimizer scheme $(k,m,h)$ with
$m > (3+\epsilon)\log_4(w+1)$ for any $\epsilon>0$ and $w=k-m+1$,
there exists a $(1-\varepsilon)$-LP MPHF for a SPSS $\mathcal{S}$
with $n=|\spectrum_k(\mathcal{S})|$
which takes
\begin{equation*}
n\cdot\frac{2}{w+1}\Big(\log_2\Big(4(w+1)^2\Big) + b + o(1)\Big) \text{ bits}
\end{equation*}
where $\varepsilon=\frac{2}{w+1}$ and $b$ is a constant larger than $\log_2(e)$.
\end{thm}

Note that the space bound in Theorem~\ref{thm:space-usage-random-minimizers-1}
decreases as $w$ grows;
for example, when $m$ is fixed and $k$ grows.
Next we show how to improve this result
using some structural properties of {\skm}s.


\subsection{Partitioned Data Structure}\label{sec:partitioning}

Property~\ref{prop:relation-size-pos} states that
$ |g|-k+1 \leq p_{g,1} \leq w$ for any {\skm} $g$.
As an immediate implication we have that
if $|g|-k+1=w$ then also $p_{g,1}=w$ (and, symmetrically,
if $p_{g,1}=1$ then $|g|=k$).
This suggests that, whenever a {\skm} contains a \emph{maximal}
number of {\kmer}s, then we can always implicitly derive that $|g|-k+1=p_{g,1}=w$.
We can thus save the space for the entries dedicated to such {\skm}s
in the arrays $L$ and $P$.
Note that the converse is not true in general, i.e.,
if $p_{g,1}=w$ it could be that $|g|-k+1 < w$.
Nonetheless, we can still save space for some entries of $P$ in this case.


Depending on the starting position of the minimizer in the \emph{first} and \emph{last} {\kmer}
of a {\skm}, we distinguish between four \emph{types} of {\skm}s (Definition~\ref{def:fl-rule}).

\begin{defn}[FL rule]\label{def:fl-rule}
Let $g$ be a {\skm}.
The first/last (FL) rule is as follows:
\begin{itemize}
    \item if $p_{g,1}=w$ and $p_{g,|g|-k+1}=1$, then $g$ is left-right-max; else
    \item if $p_{g,1}<w$ and $p_{g,|g|-k+1}=1$, then $g$ is left-max; else
    \item if $p_{g,1}=w$ and $p_{g,|g|-k+1}>1$, then $g$ is right-max; else
    \item if $p_{g,1}<w$ and $p_{g,|g|-k+1}>1$, then $g$ is non-max.
\end{itemize}
\end{defn}

See Fig.~\ref{fig:skms} for a schematic illustration.

\begin{figure}[t]
    \centering
    \includegraphics[scale=0.8,type=pdf,ext=.pdf,read=.pdf]{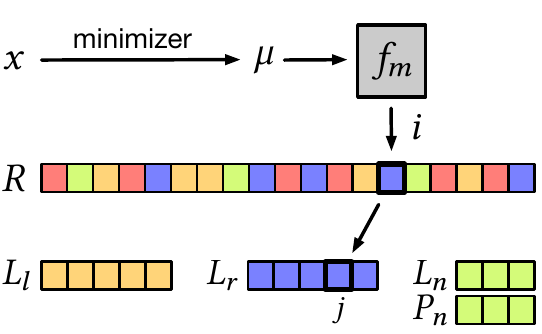}
    \vspace{-4mm}
    \caption{Partitioned data structure layout and the flow of Algorithm~\ref{alg:lookup-v2}
    for a query {\kmer} $x$, whose minimizer is $\mu$, and with $i=f_m(\mu)$.
    Different colors in $R$ are used to distinguish between the different {\skm} types.}
    \label{fig:partitioned}
    {\afterfigspace}
\end{figure}

\parag{Layout}
Based on the FL-rule above, we derive a partitioned layout as follows.
We store the type of each {\skm} in an array
$R[1..|\mathcal{M}|]$, in the order given by $f_m$.
We can now exploit this labeling of {\skm}s to improve
the space bound of Theorem~\ref{thm:space-usage-random-minimizers-1}
because:
\begin{itemize}
    \item for all left-right-max {\skm}s, we do not store $L$ nor $P$;
    \item for all left/right-max {\skm}s, we only store $L$ -- precisely, two arrays $L_l$ and $L_r$ for left- and right-max {\skm}s respectively;
    \item for all the other {\skm}s, i.e., non-max, we store both $L$ and $P$ as explained before --
    let us indicate them with $L_n$ and $P_n$ in the following.
\end{itemize}

Addressing the arrays $L_l$, $L_r$, $L_n$ and $P_n$,
can be achieved by answering $\Rank_t(i)$ queries on $R$:
the result of this query is the number of {\skm}s that have type $t$
in the prefix $R[1..i]$.
If $i=f_m(\mu)$, then we read the type of the {\skm} associated to $\mu$ as $t=R[i]$.
Then we compute $j=\Rank_t(i)$.
Depending on the type $t$, we either do not perform any array access
or access the $j$-th position of either $L_l$, or $L_r$, or $L_n$ and $P_n$
(see Algorithm~\ref{alg:lookup-v2}).

A succinct representation of $R$ that also supports $\Rank_t(i)$ and $\Access(i)$
queries is the \emph{wavelet tree}~\citep{GrossiGV03}. In our case, we only have 4
possible types, hence a 2-bit integer is sufficient to encode a type.
The wavelet tree therefore represents $R$ in
$2|\mathcal{M}|+o(|\mathcal{M}|)$ bits\footnote{The $o(|\mathcal{M}|)$ term is the redundancy
needed to accelerate the binary rank queries.
In practice, the term $o(|\mathcal{M}|)$ can be non-negligible, e.g.,
can be as high as $2\cdot(|\mathcal{M}|/4)$ bits using the Rank9 index~\citep[Sec. 3]{vigna2008broadword}, but
it is necessary for fast queries in practice (namely, $O(1)$ time).
Looking at Table 1a from~\citep{PIBIRI2021101756}, we see that the
redundancy is in between 3\% and 25\% of $2|\mathcal{M}|$.}
and supports both queries in $O(1)$ time.
The wavelet tree is also built in linear time, so the building time
of the overall data structure remains $O(n)$.
Refer to Fig.~\ref{fig:partitioned} for a pictorial representation
of this partitioned layout.

\parag{Lookup}
Algorithm~\ref{alg:lookup-v2} gives the lookup algorithm for the partitioned representation
of $f$. The complexity of the algorithm is still $O(w)$ like that of the un-partitioned counterpart,
Algorithm~\ref{alg:lookup-v1}. 
The evaluation algorithm must now distinguish between the four different types of minimizer.
On the one hand, this distinction involves an extra array access (to $R$)
and a rank query as explained above but,
on the other hand, it permits to save 2 array accesses in the left-right-max case
or 1 in the left/right-max case compared to
Algorithm~\ref{alg:lookup-v1} that \emph{always} performs 2 array accesses
(one access to $L$ and one to $P$).
Hence, the overall number of array accesses performed by Algorithm~\ref{alg:lookup-v2}
is on average the same as that of Algorithm~\ref{alg:lookup-v1} assuming
the four cases are equally likely (see next paragraph).
For this reason
we do not expect Algorithm~\ref{alg:lookup-v2} to incur
in a penalty at query time compared to Algorithm~\ref{alg:lookup-v1} despite of its more complex evaluation.

\begin{algorithm}[t]
\begin{algorithmic}[1]
\Function{$f$}{$x$}:
    \State{$(\mu,p) = \func{minimizer}(x)$}
    \State{$i=f_m(\mu)$}
    \State{$t=R[i]$}
    \State{$j=\Rank_t(i)$}
    \State{$\var{prefix}=0$, $\var{offset}=0$, $p_1=0$}

    \State{\code{switch}($t$):}
    \State{\quad\,\,\code{case} left-right-max:}
    \State{\quad\quad\,\,$\var{prefix}=0$, $\var{offset}=(j-1)w$, $p_1=w$}
    \State{\quad\quad\code{break}}
    \State{\quad\code{case} left-max:}
    \State{\quad\quad$\var{prefix}=n_{lr}$, $\var{offset}=L_l[j]$, $p_1=L_l[j+1]-L_l[j]$}
    \State{\quad\quad\code{break}}
    \State{\quad\code{case} right-max:}
    \State{\quad\quad$\var{prefix}=n_{lr} + n_l$, $\var{offset}=L_r[j]$, $p_1=w$}
    \State{\quad\quad\code{break}}
    \State{\quad\code{case} non-max:}
    \State{\quad\quad$\var{prefix}=n_{lr} + n_l + n_r$, $\var{offset}=L_n[j]$, $p_1=P_n[j]$}
    \State{\quad\quad\code{break}}

    \State{\code{return} $\var{prefix} + \var{offset} + p_1 - p$}
\EndFunction
\end{algorithmic}
\caption{Evaluation algorithm for a partitioned representation of $f$.
The quantities $n_{lr}$, $n_l$, $n_r$, and $n_n$ are, respectively,
the number of left-right-max, left-max, right-max, and non-max {\skm}s
of $\mathcal{S}$.
\label{alg:lookup-v2}}
\end{algorithm}

\parag{Compression}
Intuitively, if the fraction of left-right-max {\skm}s and
that of left/right-max {\skm}s is sufficiently high, we can save significant space
compared to the data structure in Section~\ref{sec:data-structure}
that stores both $L$ and $P$ for all minimizers.
We therefore need to compute the proportions of the different types
of {\skm}s as given by the FL rule.
For ease of notation,
let
$P_{lr}=\pr{g\,\,\txt{is left-right-max}}$,
$P_{l}=\pr{g\,\,\txt{is left-max}}$,
$P_{r}=\pr{g\,\,\txt{is right-max}}$,
$P_{n}=\pr{g\,\,\txt{is non-max}}$,
for any {\skm} $g$.

\begin{remark}
The FL rule is a partitioning rule, i.e.,
$P_{lr}+P_{l}+P_{r}+P_{n}=1$ for any {\skm}.
\end{remark}

Our objective is to derive the expression for the probabilities
$P_{lr}$, $P_{l}$, $P_{r}$, and $P_{n}$,
parametric in $k$ ({\kmer} length) and $m$ (minimizer length).
To achieve this goal we propose a simple model based on a (discrete-time) Markov chain.

\begin{figure}[t]
    \centering
    \includegraphics[scale=0.58,type=pdf,ext=.pdf,read=.pdf]{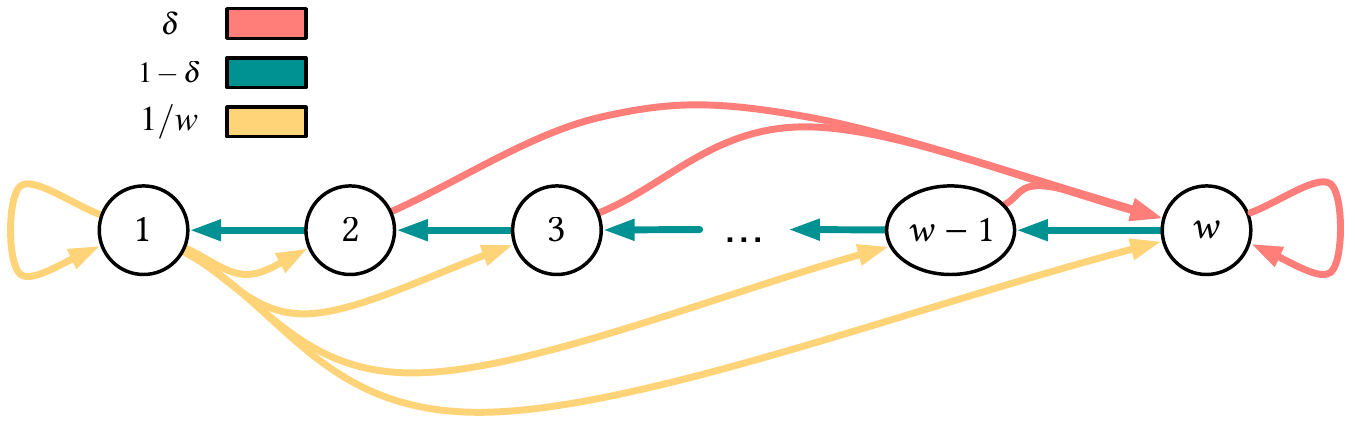}
    \caption{The chain is in state $1 \leq p \leq w$ if the minimizer starts at position $p$ in the {\kmer}.
    Different edge colors represent different probabilities.}
    \label{fig:markov}
    {\afterfigspace}
\end{figure}

Let $X : \Sigma^k \rightarrow \{1,\ldots,w\}$ be a discrete random variable,
modelling the starting position of the minimizer in a {\kmer}.
The corresponding Markov chain is illustrated in Fig.~\ref{fig:markov}.
Each state of the chain is labelled with the corresponding value assumed by $X$, i.e.,
with each value in $\{1,\ldots,w\}$.
Clearly, we have a left-right-max {\skm} if, from state $w$
we transition to state $w-1$, then to $w-2$, $\ldots$, down to state $1$.
Each state has a ``fallback'' probability to go to state $w$ which corresponds
to the event that the right-most $m$-mer (that coming next to the right) is the new minimizer.
If the chain reaches state $1$, instead, we know that we are always going to see a new
minimizer next.
If $c \in [1..u]$ is the code assigned to the current minimizer by the coding function
$h$ used by $\mu$, for some universe size $u$
(e.g, if $c$ is a 64-bit hash code, then $u=2^{64}$), the probability
for any $m$-mer to become the new minimizer is equal to $\delta=\frac{c-1}{u}$.
Vice versa, the probability of keeping the same minimizer when sliding one position
to the right, is $1-\delta$.
Whenever we change minimizer, we generate a new code $c$
and, hence, the probability $\delta$ changes with every formed {\skm}.
Nonetheless, the following Theorem shows that the probabilities $P_{lr}$, $P_{l}$, $P_{r}$,
and $P_{n}$, do not depend on $\delta$.

\begin{thm}\label{thm:skms}
For any random minimizer scheme $(k,m,h)$ we have
\begin{align*}
    P_{lr} &= \pr{g\,\,\txt{is left-right-max}} = W^2 + 1/w \\
    P_{l}  &= \pr{g\,\,\txt{is left-max}} = W(1-W) \\
    P_{r}  &= \pr{g\,\,\txt{is right-max}} = W(1-W) \\
    P_{n}  &= \pr{g\,\,\txt{is non-max}} = W^2
\end{align*}
where $W=\frac{1}{2}\cdot(1-\frac{1}{w})$ and $w=k-m+1$.
\end{thm}

We give the following lemma to prove Theorem~\ref{thm:skms}.
(When we write ``first''/``last'' {\kmer} we are going to silently
assume ``of a {\skm}''.)

\begin{lem}\label{lem:P1}
$\pr{X=1}=\frac{1}{2}$ and $\pr{X=w}=\frac{1}{2}\cdot(1+\frac{1}{w})$.
\end{lem}

\begin{proof}
First note that
\begin{equation}\label{eq:0}
\pr{X=p \,\, \txt{in the \textbf{first} {\kmer}}} = \pr{X=1} \cdot \frac{1}{w},
\end{equation}
for any $1 \leq p \leq w-1$.
Then we have the following equivalences.
\begin{align}\label{eq:1}
& \sum_{p=1}^w \pr{X=p \,\, \txt{in the \textbf{first} {\kmer}}} = 1 \nonumber \iff \\
& \pr{X=w} + \sum_{p=1}^{w-1} \pr{X=p \,\, \txt{in the \textbf{first} {\kmer}}} = 1  \nonumber \iff \\
& \pr{X=w} + \pr{X=1}\cdot\Big(1-\frac{1}{w}\Big) = 1 \,\, [\txt{for Equation~\ref{eq:0}}].
\end{align}
Now note that
\begin{equation}\label{eq:Xw}
\pr{X=w} = P_{lr} + P_{r}
\end{equation}
because the starting position of the minimizer
of the \textbf{first} {\kmer} of any left-right-max and of any right-max {\skm} is $w$.
In a similar way, we have that
\begin{align}\label{eq:X1}
& \pr{X=1 \,\, \txt{in the \textbf{last} {\kmer}}} = \nonumber \\
& \pr{X=1} + \pr{X=1 \,\, \txt{in the \textbf{first} {\kmer}}} = \nonumber \\
& \pr{X=1}\cdot\Big(1+\frac{1}{w}\Big) \,\, \txt{[for Equation~\ref{eq:0} with } p=1] \nonumber \\
& = P_{lr} + P_{l}
\end{align}
because the starting position of the minimizer of the \textbf{last} {\kmer}
of any left-right-max and of any left-max {\skm} is 1.
Now note that $P_l=P_r$ because:
\begin{align*}
    P_l &= \pr{X=w \,\, \txt{in \textbf{first} {\kmer}}} \cdot \pr{X \neq 1 \,\, \txt{in \textbf{last} {\kmer}}} = \\
    &= (1-\pr{X \neq w \,\, \txt{in \textbf{first} {\kmer}}}) \cdot (1-\pr{X=1 \,\, \txt{in \textbf{last} {\kmer}}}) = \\
    &= \Big( 1-\pr{X=1}\cdot\Big(1-\frac{1}{w}\Big) \Big) \cdot \Big( 1-\pr{X=1}\cdot\Big(1+\frac{1}{w}\Big) \Big) = \\
    &= (\pr{X=1})^2 \cdot \Big(1-\frac{1}{w^2}\Big), \,\, \txt{and similarly} \\
    P_r &= \pr{X \neq w \,\, \txt{in \textbf{first} {\kmer}}} \cdot \pr{X=1 \,\, \txt{in \textbf{last} {\kmer}}} = \\
    &= \pr{X=1}\cdot\Big(1-\frac{1}{w}\Big) \cdot \pr{X=1}\cdot\Big(1+\frac{1}{w}\Big) = \\
    &= (\pr{X=1})^2 \cdot \Big(1-\frac{1}{w^2}\Big).
\end{align*}
From equation $P_l=P_r$, we have $P_{lr} + P_{l} = P_{lr} + P_{r}$ which,
using Equation~\ref{eq:Xw} and Equation~\ref{eq:X1}, yields
    $\pr{X=w} = \pr{X=1}\cdot(1+\frac{1}{w})$.
The Lemma follows by using the latter equation into Equation~\ref{eq:1}.
\qedsymbol
\end{proof}

Now we prove Theorem~\ref{thm:skms}.

\begin{proof}
Since the FL rule induces a partition:
\begin{align}\label{eq:Plr}
    & P_{lr}+P_r+P_l+P_n = 1 \iff \nonumber \\
    & P_{lr}+P_{lr}+P_r+P_l+P_n = 1+P_{lr} \nonumber \\
    & [\txt{adding} \,\, P_{lr} \,\, \txt{to both sides}] \iff \nonumber \\
    & 2\pr{X=w} + P_n = 1+P_{lr} \nonumber \\
    & [\txt{knowing that} \,\, P_{lr}+P_r=P_{lr}+P_l=\pr{X=w}] \iff \nonumber \\
    & P_{lr} = P_n + \frac{1}{w} \,\, [\txt{for Lemma~\ref{lem:P1}}].
\end{align}
Again exploiting the fact that $P_{lr}+P_r=P_{lr}+P_l=\pr{X=w}$, we also have
\begin{equation}\label{eq:Pl}
    P_l=P_r=\pr{X=w}-P_{lr}=\frac{1}{2}\cdot\Big(1+\frac{1}{w}\Big) - P_n - \frac{1}{w}.
\end{equation}
We have therefore to compute $P_n$ to also determine $P_{lr}$, $P_l$, and $P_r$.
\begin{align}\label{eq:Pn}
    & P_n = \pr{X \neq w \,\, \txt{in \textbf{first} {\kmer}}} \cdot \pr{X \neq 1 \,\, \txt{in \textbf{last} {\kmer}}} = \\
    & \pr{X=1}\cdot\Big(1-\frac{1}{w}\Big) \cdot (1-\pr{X=1 \,\, \txt{in \textbf{last} {\kmer}}}) = \nonumber \\
    & \pr{X=1}\cdot\Big(1-\frac{1}{w}\Big) \cdot \Big(1-\pr{X=1}\cdot\Big(1+\frac{1}{w}\Big)\Big) = \nonumber \\
    & \Big(\frac{1}{2}\cdot\Big(1-\frac{1}{w}\Big)\Big)^2 \,\,
        \txt{[for Lemma~\ref{lem:P1}]}. \nonumber
\end{align}
Now letting $W=\frac{1}{2}\cdot(1-\frac{1}{w})$ and substituting $P_n = W^2$ (Equation~\ref{eq:Pn})
into Equation~\ref{eq:Plr} and~\ref{eq:Pl}, the Theorem follows.
\qedsymbol
\end{proof}

In Table~\ref{tab:probabilities} we report the probabilities
$P_{lr}$, $P_l$, $P_r$, and $P_n$ computed using Theorem~\ref{thm:skms}
for some representative combinations of $k$ and $m$
(these combinations are some of those used in the experiments
of Section~\ref{sec:experiments}; see also Table~\ref{tab:m}).
For comparison, we also report the probabilities measured over the whole
human genome.
We see that the probabilities computed with the formulas in Theorem~\ref{thm:skms}
accurately model the empirical probabilities.

The net result is that, for sufficiently large $w$, the probabilities
in Theorem~\ref{thm:skms} are all approximately equal to 1/4,
so that we have $\approx\frac{n}{2(w+1)}$ {\skm}s of each type.
This also implies that the choice of 2-bit codes for the symbols of $R$
is essentially optimal.
Under this condition, we give the following theorem
(see the Supplementary material for the proof).

\begin{table}[t]
\centering
\caption{Computed (compt.) probabilities with Theorem~\ref{thm:skms} vs. measured (measr.)
using the whole human genome for three representative $(k,m)$ configurations.
}
\vspace{-4mm}
\scalebox{0.9}{

\begin{tabular}{
l c
cc c cc c cc
}

\toprule

&& \multicolumn{2}{c}{$k=31,m=21$}
&& \multicolumn{2}{c}{$k=47,m=26$}
&& \multicolumn{2}{c}{$k=63,m=28$}
\\
\cmidrule(lr){3-4}
\cmidrule(lr){6-7}
\cmidrule(lr){9-10}
&& compt. & measr.
&& compt. & measr.
&& compt. & measr. \\

\midrule

$P_{lr}$
&& 0.297 & 0.281
&& 0.273 & 0.264
&& 0.264 & 0.257
\\

$P_{l}$
&& 0.248 & 0.261
&& 0.249 & 0.256
&& 0.250 & 0.254
\\

$P_{r}$
&& 0.248 & 0.261
&& 0.249 & 0.256
&& 0.250 & 0.254
\\

$P_{n}$
&& 0.207 & 0.197
&& 0.228 & 0.224
&& 0.236 & 0.235
\\


\bottomrule
\end{tabular}



}
\label{tab:probabilities}
\end{table}

\begin{thm}\label{thm:space-usage-random-minimizers-2}
Given a random minimizer scheme $(k,m,h)$ with
$m > (3+\epsilon)\log_4(w+1)$ for any $\epsilon>0$ and $w=k-m+1$,
there exists a $(1-\varepsilon)$-LP MPHF for a SPSS $\mathcal{S}$
with $n=|\spectrum_k(\mathcal{S})|$
which takes
\begin{equation*}
n\cdot\frac{2}{w+1}\Big(\log_2\Big(\frac{16 \cdot 2^{1/4}}{3}(w+1)\Big) + b + o(1)\Big)
\text{ bits}
\end{equation*}
where $\varepsilon=\frac{2}{w+1}$ and $b$ is a constant larger than $\log_2(e)$.
\end{thm}


\subsection{Ambiguous Minimizers}\label{sec:ambiguous-min}

Let $G_{\mu}$ be the set of {\skm}s whose minimizer is $\mu$.
The rank computation in Equation~\ref{eq:local-rank} can be used
as long as $|G_{\mu}|=1$, i.e., whenever one single {\skm} $g$ has minimizer $\mu$
and, thus, the single $p_{g,1}$ unequivocally displace
all the {\kmer}s $x_{g,1},\ldots,x_{g,|g|-k+1}$.
When $|G_{\mu}|>1$ we say that the minimizer $\mu$ is ``ambiguous''.
It is a known fact that the number of such minimizers is very small
for a sufficiently long minimizer length $m$~\citep{pibiri2022sshash,JainRZCWKP20,chikhi2014representation},
and the number decreases for growing $m$.
For example, on the datasets used in Section~\ref{sec:experiments},
the fraction of ambiguous minimizers is in between 1\% and 4\%.
However, they must be dealt with in some way.

Let $\xi$ be the fraction of {\kmer}s whose minimizers
are ambiguous. Our strategy is to build a fallback MPHF
for these {\kmer}s. This function adds $\xi\cdot b$ bits/{\kmer}
on top of the space of Theorem~\ref{thm:space-usage-random-minimizers-1}
and Theorem~\ref{thm:space-usage-random-minimizers-2}, where $b > \log_2(e)$ is the number of
bits per key spent by a MPHF of choice.
The fallback MPHF makes our functions $(1-\varepsilon+\xi)$-locality-preserving.

To detect ambiguous minimizers, one obvious option would be
to explicitly use an extra 1-bit code per minimizer.
This would however result in a waste of 1 bit per minimizer for most of them
since we expect to have a small percentage of ambiguous minimizers.
To avoid these problems, we use the following trick.
Suppose $\mu$ is an ambiguous minimizer. We initially pretend that $\mu$
is not ambiguous.
For the un-partitioned data structure from Section~\ref{sec:data-structure},
we set $L[f_m(\mu)]=0$. A {\skm} of size 0 is clearly not possible,
thus we use the value 0 to indicate that $\mu$ is actually ambiguous.
We do the same for the partitioned data structure from Section~\ref{sec:partitioning}:
in this case we set $L_r[f_m(\mu)]=0$ pretending the type of $\mu$
is right-max (but we could have also used the type left-max or non-max).
To sum up, with just an extra check on the {\skm} size
we know if the query {\kmer} must be looked-up in the fallback MPHF or not.

We leave the exploration of alternative strategies to handle ambiguous
minimizers to future work.
For example, one can imagine a recursive data structure
where, similarly to~\citep{shibuya2022space},
each level is an instance of the construction
with different minimizer lengths:
if level $i$ has minimizer length $m_i$,
then level $i+1$ is built with length $m_{i+1} > m_i$
over the {\kmer}s whose minimizers
are ambiguous at level $i$.

\begin{figure*}[t]
    \centering
    \includegraphics[scale=0.85,type=pdf,ext=.pdf,read=.pdf]{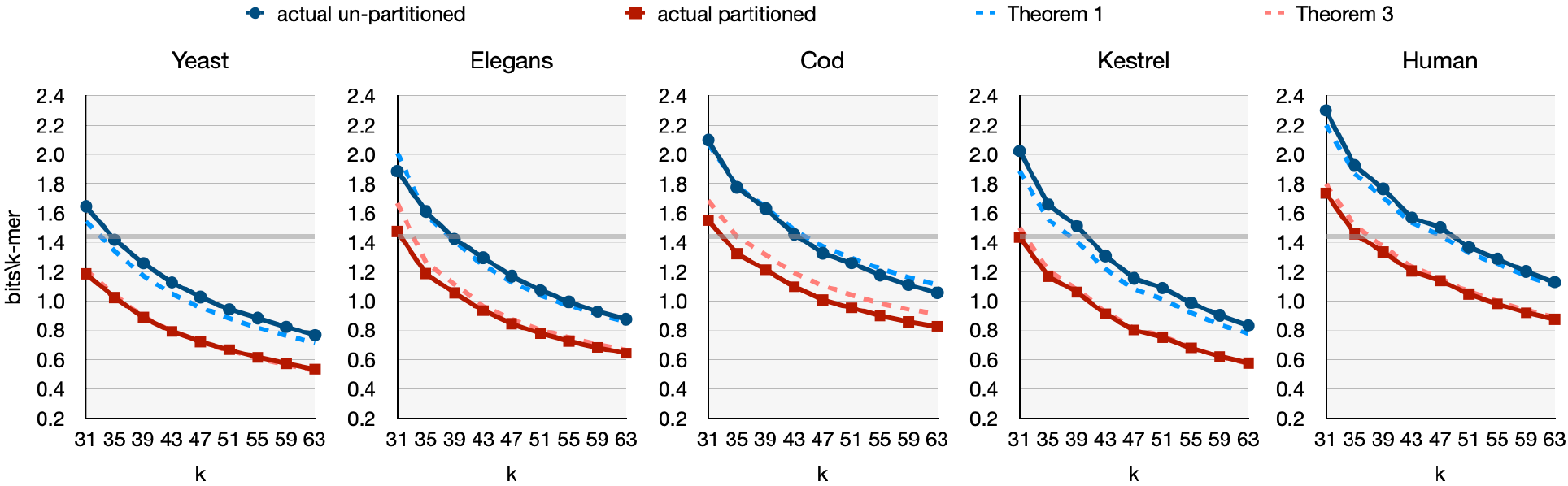}
    \vspace{-5mm}
    \caption{Space in average bits/{\kmer} for LPHash by varying $k$,
    for both un-partitioned and partitioned data structures.
    The flat solid line at $\log_2(e) = 1.442$ bits/{\kmer} indicates the classic MPHF lower-bound.
    Lastly, the dashed lines corresponds to the space bounds computed using Theorem~\ref{thm:space-usage-random-minimizers-1}
    and Theorem~\ref{thm:space-usage-random-minimizers-2} with $b=2.5$ and including the space for the fallback MPHF.
    }
    \label{fig:space}
    \raggedbottom
\end{figure*}

\section{Experiments}\label{sec:experiments}

In this section we report on the experiments conducted to asses the
practical performance of the method presented in Section~\ref{sec:upper-bound},
which we refer to as LPHash in the following.
Our implementation of the method is written in C++
and available at \url{https://github.com/jermp/lphash}.

\parag{Implementation Details}
We report here the major implementation details for LPHash.
The arrays $L$ and $P$ are compressed with Elias-Fano~\citep{Fano71,Elias74}
to exploit its constant-time random access
(see also~\citep[Sec. 3.4]{PibiriV21} for an explanation of such compressed encoding).
Both the function $f_m$ and the fallback MPHF are implemented
with PTHash using parameters $(\textup{D-D},\alpha=0.94,c=3.0)$, unless otherwise
specified. Under this configuration the space taken by a PTHash MPHF is $2.3-2.5$ bits/key.

We do not compress the bit-vectors in the wavelet tree
and we add constant-time support for rank
queries using the Rank9 index~\citep{jacobson1989space,vigna2008broadword}.
The Rank9 index adds $25\%$ more space at each level of the wavelet tree,
making the wavelet tree to take 2.5 bits per element in practice.
Therefore, we estimate the little-Oh factor in Theorem~\ref{thm:space-usage-random-minimizers-1}
and Theorem~\ref{thm:space-usage-random-minimizers-2}
to be 0.5.

\begin{table}[t]
\centering
\caption{Minimizer length $m$ by varying $k$ on the different datasets.
}
\scalebox{0.9}{

\begin{tabular}{
l c c
ccccccccc
}

\toprule

& $k \rightarrow$
& 31 & 35 & 39 & 43 & 47 & 51 & 55 & 59 & 63
\\

\midrule

{\yeast} & &  15 & 15 & 16 & 16 & 16 & 16 & 18 & 18 & 18 \\
{\celeg} & &  16 & 18 & 18 & 20 & 20 & 20 & 20 & 20 & 20 \\
{\cod}   & &  20 & 20 & 22 & 22 & 22 & 24 & 24 & 24 & 24 \\
{\kes}   & &  20 & 20 & 22 & 22 & 22 & 24 & 24 & 24 & 24 \\
{\human} & &  21 & 21 & 23 & 23 & 26 & 26 & 28 & 28 & 28 \\

\bottomrule
\end{tabular}}
\label{tab:m}
\raggedbottom
\end{table}

\parag{Competitors}
We compare the space usage, query time, and building time of LPHash
against PTHash~\citep{abs-2106-02350,PibiriT21}, the fastest MPHF in the literature,
and the popular BBHash~\citep{limasset2017fast}.
Both competitors are also written in C++.
Following the recommendations of the respective authors, we tested two example
configurations each:
\begin{itemize}
    \item PTHash-v1, with parameters $(\textup{D-D},\alpha=0.94,c=5.0)$;
    \item PTHash-v2, with parameters $(\textup{EF},\alpha=0.99,c=5.0)$;
    \item BBHash-v1, with parameter $\gamma=2$;
    \item BBHash-v2, with parameter $\gamma=1$;
\end{itemize}
We point the reader to the respective papers for an explanation of such parameters;
we just report that they offer a trade-off between space, query efficiency, and building time
as also apparent in the following experiments.

\parag{Testing Machine}
The experiments were executed on a machine equipped with
a Intel i9-9900K CPU (clocked at 3.60GHz), 64 GB of RAM,
and running the Linux 5.13.0 operating system.
The whole code (LPHash and competitors) was compiled with gcc 11.2.0,
using the flags \texttt{-O3} and \texttt{-march=native}.

\parag{Datasets}
We use datasets of increasing size in terms of number of distinct {\kmer}s;
namely, the whole-genomes of:
\emph{Saccharomyces Cerevisiae} ({\yeast}, $11.6 \times 10^6$ {\kmer}s),
\emph{Caenorhabditis Elegans} ({\celeg}, $96.5 \times 10^6$ {\kmer}s),
\emph{Gadus Morhua} ({\cod}, $0.56 \times 10^9$ {\kmer}s),
\emph{Falco Tinnunculus} ({\kes}, $1.16 \times 10^9$ {\kmer}s),
and \emph{Homo Sapiens} ({\human}, $2.77 \times 10^9$ {\kmer}s).
For each dataset, we obtain the corresponding SPSS
by first building the compacted de Bruijn graph using BCALM2~\citep{chikhi2016compacting},
then running the UST algorithm~\citep{rahman2020representation}.
At our code repository 
we provide detailed instructions on how to
prepare the datasets for indexing.
Also, all processed datasets
are available at \url{https://zenodo.org/record/7239205} already in processed form
so that it is easy to reproduce our results.


\begin{table}[t]
\centering
\caption{Space in average bits/{\kmer} for PTHash and BBHash.\tablefootnote{\red{The numbers reported in Table~\ref{tab:space-competitors} were taken for $k=63$ although the
avg. bits/{\kmer} for PTHash and BBHash does not depend on $k$.}}
\red{As reference points, we also report the bits/{\kmer} for partitioned LPHash for three representative values of $k$ (see also Fig.~\ref{fig:space}).}}
\scalebox{0.9}{

\begin{tabular}{
l c
c c c c c c c c c c
}

\toprule

Method
&
$k$
&& {\yeast}
&& {\celeg}
&& {\cod}
&& {\kes}
&& {\human}
\\

\midrule

\multirow{3}{*}{{LPHash}}

& 31
&& \red{1.18}
&& \red{1.47}
&& \red{1.55}
&& \red{1.43}
&& \red{1.74}
\\

& 47
&& \red{0.72}
&& \red{0.85}
&& \red{1.01}
&& \red{0.82}
&& \red{1.14}
\\

& 63
&& \red{0.53}
&& \red{0.64}
&& \red{0.83}
&& \red{0.58}
&& \red{0.87}
\\

\cmidrule(lr){1-12}

PTHash-v1
&
&&  2.76
&&  2.68
&&  2.65
&&  2.58
&&  2.65
\\

PTHash-v2
&
&&  2.20
&&  2.13
&&  2.09
&&  2.06
&&  2.04
\\

BBHash-v1
&
&& 3.71
&& 3.71
&& 3.71
&& 3.71
&& 3.71
\\

BBHash-v2
&
&& 3.06
&& 3.06
&& 3.06
&& 3.06
&& 3.06
\\

\bottomrule
\end{tabular}}
\label{tab:space-competitors}
\end{table}

\subsection{Space Effectiveness}
To build an instance of LPHash for a given $k$, we have to choose a suitable value
of minimizer length ($m$).
A suitable value of $m$ should clearly be not too small (otherwise,
most minimizers will appear many times), nor too large (otherwise, the space of $f_m$
will be too large as well).
In general, a good value for $m$ can be chosen around $\log_4(N)$ where
$N$ is the cumulative length of the strings in the input SPSS.
Remember from our discussion in Section~\ref{sec:ambiguous-min}
that the fraction of ambiguous minimizers decreases for growing $m$.
Therefore, testing LPHash for growing values of $k$ allows us to
progressively increase $m$, starting from $m=\log_4(N)$, while keeping $w=k-m+1$
sufficiently large and reducing the fraction of ambiguous minimizers as well.
Following this principle, for each combination of $k$ and dataset,
we choose $m$ as reported in Table~\ref{tab:m}.

Fig.~\ref{fig:space} shows the space of LPHash in average bits/{\kmer},
by varying $k$ from 31 to 63 with a step of 4,
for both un-partitioned and partitioned data structures.
We report the actual space usage achieved by the implementation against
the space bounds computed using Theorem~\ref{thm:space-usage-random-minimizers-1}
(un-partitioned)
and Theorem~\ref{thm:space-usage-random-minimizers-2} (partitioned)
for $b=2.5$. The $b$ parameter models the number of bits per key spent
by a MPHF of choice for the representation of the minimizer MPHF and the fallback MPHF.
(For all datasets we use $c=3.0$ for the PTHash $f_m$ and fallback, except
on the largest {\human} where we use $c=5.0$ to lower construction time
at the expense of a larger space usage.)

\begin{table*}[t]
\centering
\caption{Query time in average nanoseconds per {\kmer}. 
}
\scalebox{0.9}{

\begin{tabular}{
l ccc
cc c cc c cc c cc c cc
}

\toprule

\multirow{2}{*}{Method}
&& \multirow{2}{*}{$k$}
&& \multicolumn{2}{c}{\yeast}
&& \multicolumn{2}{c}{\celeg}
&& \multicolumn{2}{c}{\cod}
&& \multicolumn{2}{c}{\kes}
&& \multicolumn{2}{c}{\human}
\\
\cmidrule(lr){5-6}
\cmidrule(lr){8-9}
\cmidrule(lr){11-12}
\cmidrule(lr){14-15}
\cmidrule(lr){17-18}
&&
&& stream & random
&& stream & random
&& stream & random
&& stream & random
&& stream & random \\

\midrule

\multirow{9}{*}{{LPHash}}

&&
31
&&  29 & 110
&&  40 & 118
&&  79 & 144
&&  84 & 145
&& 107 & 162
\\

&&
35
&& 28 & 125
&& 35 & 124
&& 65 & 147
&& 69 & 149
&& 90 & 166
\\

&&
39
&& 27 & 130
&& 32 & 131
&& 60 & 149
&& 63 & 153
&& 82 & 166
\\

&&
43
&& 25 & 137
&& 30 & 135
&& 52 & 152
&& 54 & 155
&& 73 & 169
\\

&&
47
&& 24 & 145
&& 28 & 143
&& 47 & 155
&& 49 & 159
&& 69 & 172
\\

&&
51
&& 24 & 152
&& 28 & 150
&& 45 & 159
&& 46 & 162
&& 63 & 174
\\

&&
55
&& 23 & 157
&& 26 & 157
&& 41 & 165
&& 42 & 167
&& 59 & 176
\\

&&
59
&& 23 & 165
&& 25 & 165
&& 39 & 171
&& 39 & 173
&& 57 & 182
\\

&&
63
&& 22 & 174
&& 24 & 172
&& 37 & 180
&& 37 & 179
&& 53 & 188
\\

\midrule

PTHash-v1
&&
&& \multicolumn{2}{c}{24}
&& \multicolumn{2}{c}{46}
&& \multicolumn{2}{c}{67}
&& \multicolumn{2}{c}{72}
&& \multicolumn{2}{c}{72}
\\

PTHash-v2
&&
&& \multicolumn{2}{c}{38}
&& \multicolumn{2}{c}{64}
&& \multicolumn{2}{c}{130}
&& \multicolumn{2}{c}{155}
&& \multicolumn{2}{c}{175}
\\

BBHash-v1
&&
&& \multicolumn{2}{c}{42}
&& \multicolumn{2}{c}{118}
&& \multicolumn{2}{c}{170}
&& \multicolumn{2}{c}{175}
&& \multicolumn{2}{c}{175}
\\

BBHash-v2
&&
&& \multicolumn{2}{c}{42}
&& \multicolumn{2}{c}{125}
&& \multicolumn{2}{c}{180}
&& \multicolumn{2}{c}{190}
&& \multicolumn{2}{c}{190}
\\

\bottomrule
\end{tabular}}
\label{tab:query-time}
\end{table*}

We make the following observations.

\begin{itemize}

\item
The space bounds computed with
Theorem~\ref{thm:space-usage-random-minimizers-1}
and Theorem~\ref{thm:space-usage-random-minimizers-2}
are very similar to the actual space usage of LPHash,
thus confirming the correctness and accuracy of our analysis in Section~\ref{sec:upper-bound}.

\item As expected, the space of LPHash lowers for increasing $k$ and the partitioned data structure
is always considerably smaller than the un-partitioned counterpart.

\item We report the space taken by the tested competitive configurations in Table~\ref{tab:space-competitors}.
Comparing the space values in Table~\ref{tab:space-competitors} with those in
Fig.~\ref{fig:space},
the net result is that the space of LPHash is much lower than that
of the classic MPHFs traditionally used in the prior literature and in practice.

To make a concrete example, partitioned LPHash for $k=63$
achieves 0.53, 0.64, 0.83, 0.58, and 0.87 bits/{\kmer} on
{\yeast}, {\celeg}, {\cod}, {\kes}, and {\human} respectively.
These values are $5.1\times$, $4.1\times$, $3.2\times$, $4.4\times$, and $3\times$
smaller than the those achieved by PTHash-v1 
(and even smaller when compared to BBHash).
Even compared to the most succinct configuration, PTHash-v2 (around 2 bits/{\kmer}),
LPHash still retains $2.3-3.7\times$ better space.

We remark that, however, PTHash and BBHash are ``general-purpose'' MPHFs that
can work with arbitrary keys, whereas
the applicability of LPHash is restricted to spectrum-preserving string sets.

\end{itemize}

\subsection{Query Time}
Table~\ref{tab:query-time} reports the query time for LPHash in comparison
to PTHash and BBHash. Timings were collected using a single core of the processor.
We query all {\kmer}s read from the Human chromosome 13,
for a total of $\approx100\times 10^6$ queries.
First of all, we report that query timings for un-partitioned and
partitioned LPHash are the same,
so we do not distinguish between the two data structures
in Table~\ref{tab:query-time}.
This meets our expectation regarding the average number of array accesses
that the two query algorithms perform as explained in Section~\ref{sec:partitioning}.


We distinguish between streaming and random queries (lookups) for LPHash.
Given a query string $Q$, we query for each {\kmer} read \emph{consecutively}
from $Q$, that is, for $Q[1..k]$, $Q[2..k+1]$, $Q[3..k+2]$, etc.
We refer to the this query modality as streaming; anything else
different from streaming is a random lookup
(i.e., ``random'' here means ``without locality'').
LPHash is optimized for streaming lookup queries,
whereas PTHash and BBHash do not benefit from any specific query order.
In fact, the locality-preserving nature of LPHash makes the calculation
of hashes for consecutive {\kmer}s very cheap, as consecutive {\kmer}s
are likely to be part of the same {\skm}.

Considering the result in Table~\ref{tab:query-time}, we see that
LPHash's streaming query time is in fact much smaller
than random query time.
Both timings are sensitive to the growth of $k$:
while the streaming one slightly decreases for the better locality,
the random one increases instead, for the more expensive hash calculations.


LPHash is as fast as PTHash-v1 (fastest configuration) for streaming queries on
the smaller {\yeast} dataset, but actually up to $1.4-2\times$ faster on the larger
datasets {\celeg}, {\cod}, {\kes}, and {\human}.
Instead, it is up to $4\times$ faster than PTHash-v2.
We stress that this is a remarkable result given that PTHash is the fastest MPHF
in the literature, being $2-6\times$ faster than other methods.
Compared to BBHash, LPHash is $2\times$ faster on {\yeast}
and up to $4-5\times$ faster on the larger datasets.

Random lookup time is, instead, slower for LPHash compared to
PTHash: this is expected because
the evaluation of LPHash is more complex (it involves computing the minimizer,
accessing several arrays, and computing a rank using a wavelet tree).
However, we do not regard this as a serious limitation since,
as we already motivated, the streaming query modality is the one used
in Bioinformatics tasks involving {\kmer}s~\citep{almodaresi2018space,bingmann2019cobs,blight,findere,pibiri2022sshash}.
We also observe that the slowdown is more evident on the smaller datasets
while it tends to diminish on the larger ones.
Except for the smaller {\yeast} dataset, the random lookup time of LPHash
is competitive with that of BBHash or better.

\begin{table}[t]
\centering
\caption{Total building time, including the time to read the input and serialize
the data structure on disk.
All constructions were run with 4 processing threads.}
\scalebox{0.9}{

\begin{tabular}{
l c
c c c c c c c c r
}

\toprule

\multirow{2}{*}{Method}
&& {\yeast}
&& {\celeg}
&& {\cod}
&& {\kes}
&& {\human}
\\
\cmidrule(lr){2-11}
&& mm:ss
&& mm:ss
&& mm:ss
&& mm:ss
&& mm:ss \\

\midrule

LPHash
&& 00:01
&& 00:15
&& 05:30
&& 03:50
&& 07:25
\\

\cmidrule(lr){1-11}

PTHash-v1
&&  00:03
&&  00:29
&&  07:37
&&  20:34
&&  63:30
\\
PTHash-v2
&&   00:03
&&   00:46
&&   14:15
&&   40:00
&&  124:00
\\

BBHash-v1
&& 00:01
&& 00:07
&& 00:48
&& 01:40
&& 04:13
\\
BBHash-v2
&& 00:01
&& 00:08
&& 01:05
&& 02:22
&& 07:50
\\

\bottomrule
\end{tabular}

}
\label{tab:construction-time}
\end{table}

\subsection{Building Time}
We now consider building time which is reported
in Table~\ref{tab:construction-time}.
Both LPHash and PTHash were built limiting to 8GB the maximum amount of RAM to use
before resorting to external memory.
(There is no such capability in the BBHash implementation so BBHash took
more RAM at building time than the other two constructions.)

The building time for un-partitioned and partitioned LPHash is the same.
LPHash is competitive with the fastest BBHash
and significantly faster than PTHash on the larger datasets.
Specifically, it is faster than PTHash over the entire set of {\kmer}s
since it builds two smaller PTHash functions ($f_m$ and fallback).
The slowdown seen for {\cod} is due to the larger fallback MPHF,
which is built with PTHash under a strict configuration ($c=3.0$) that privileges
space effectiveness (and query efficiency) rather than building time.
One could in principle use BBHash instead of PTHash for the fallback function,
hence trading space for better building time.
For example, recall that we use $c=5.0$ on {\human} for this reason.

\vspace{-5mm}

\section{Conclusion and Future Work}\label{sec:conclusion}

In this paper,
we initiate the study of locality-preserving minimal perfect hash
functions for {\kmer}s. We propose a
construction, named LPHash, that achieves very compact space
by exploiting the fact that consecutive {\kmer}s share overlaps
of $k-1$ symbols. This allows LPHash to actually break the theoretical $\log_2(e)$ bit/key
barrier for minimal perfect hash functions.

We show that a concrete implementation of the method is practical as well.
Before this paper, one used to build a BBHash function over the {\kmer}s
and spend (approximately) 3 bits/{\kmer} and 100-200 nanoseconds per lookup.
This work shows that it is possible to do significantly better than this
when the {\kmer}s come from a spectrum-preserving string set:
for example, less than 0.6-0.9 bits/{\kmer} and 30-60 nanoseconds per lookup.
Our code is open-source.

As future work,
we plan to further engineer the current implementation
to accelerate construction and streaming queries.
Other strategies for sampling the strings
could be used other than random minimizers~\citep{Frith2022.08.18.504476};
for example,
the \emph{Miniception}~\citep{zheng2020improved} achieving
$\varepsilon=\frac{1.67}{w} + o(1/w)$.
Evaluating the impact of such different sampling schemes
is a promising avenue for future research.
Lastly, we also plan to investigate other strategies
for handling the ambiguous minimizers.
A better strategy is likely to lead
to improved space effectiveness and faster construction.

\vspace{-5mm}
\section*{Acknowledgments}
The first author wishes to thank Piotr Beling for useful comments on an early draft of the paper.

\vspace{-5mm}
\section*{Funding}
This work was partially supported by the project MobiDataLab
(EU H2020 RIA, grant agreement N\textsuperscript{\b{o}}101006879)
and by the French ANR AGATE (ANR-21-CE45-0012).

\vspace{-5mm}
\bibliographystyle{plainnat}
\bibliography{bibliography}

\end{document}